\documentclass[epsf]{emulateapj}
\usepackage{epsfig}
\slugcomment{Astrophysical Journal Letters, in press:
received 9 December 2011, accepted 31 January 2012}

\shorttitle{}
\shortauthors{}

\newcommand{\lta}{\lesssim}
\newcommand{\gta}{\gtrsim}
\newcommand{\kpc}{\>{\rm kpc}}

\usepackage{subfigure}

\begin{document}

\title{Dwarfs gobbling dwarfs: a stellar tidal stream around NGC~4449\\
and hierarchical galaxy formation on small scales}

\author{David Mart\'{i}nez-Delgado\altaffilmark{1,13},
Aaron J.  Romanowsky\altaffilmark{2}, 
R. Jay Gabany\altaffilmark{3}, 
Francesca Annibali\altaffilmark{4},\\
Jacob A. Arnold\altaffilmark{2},
J\"urgen Fliri\altaffilmark{5,6}, 
Stefano Zibetti\altaffilmark{7}, 
Roeland P.~van der Marel\altaffilmark{8},\\ 
Hans-Walter Rix\altaffilmark{1},
Taylor S. Chonis\altaffilmark{9}, 
Julio A. Carballo-Bello\altaffilmark{10}, 
Alessandra Aloisi\altaffilmark{8},\\
Andrea V.  Macci\`o\altaffilmark{1},
J. Gallego-Laborda\altaffilmark{11},
Jean P. Brodie\altaffilmark{2}, 
Michael R.  Merrifield\altaffilmark{12}}

\altaffiltext{1} {Max-Planck-Institut fur Astronomy, Heidelberg, Germany}
\altaffiltext{2} {University of California Observatories, 1156 High Street, Santa Cruz, CA
  95064, USA}
\altaffiltext{3} {Black Bird Observatory, Mayhill, New Mexico, USA}

\altaffiltext{4} {Osservatorio Astronomico di Bologna, INAF, Via Ranzani
       1, I-40127 Bologna, Italy}
\altaffiltext{5} {LERMA, CNRS UMR 8112, Observatoire de Paris, 61 Avenue de l'Observatoire,
75014 Paris, France}

\altaffiltext{6} {GEPI, CNRS UMR 8111, Observatoire de Paris, 5 Place Jules Janssen, 92195
Meudon, France}
\altaffiltext{7} {Dark Cosmology Centre, Niels Bohr Institute - University of
Copenhagen, Juliane Maries Vej 30, DK-2100 Copenhagen, Denmark}
\altaffiltext{8} {Space Telescope Science Institute, 3700 San Martin Drive, 
       Baltimore, MD 21218 }
\altaffiltext{9}{Department of Astronomy, University of Texas at Austin,
Texas, USA}
\altaffiltext{10} {Instituto de Astrofisica de Canarias, Tenerife, Spain}
\altaffiltext{11} {Fosca Nit Observatory, Montsec Astronomical Park, Ager, Spain}
\altaffiltext{12} {School of Physics and Astronomy, University of Nottingham,
  University Park, Nottingham NG7 2RD, England}
\altaffiltext{13} {Alexander von Humboldt Fellow for Advanced Research}


\begin{abstract}
A candidate diffuse stellar substructure was previously reported in the halo of
the nearby dwarf starburst galaxy NGC~4449 by Karachentsev et al.
We map and analyze this feature using a unique combination of
deep integrated-light images from the Black Bird 
0.5-meter telescope, and high-resolution wide-field images from
the 8-meter Subaru telescope, which resolve the nebulosity into a stream of
red giant branch stars, and confirm its physical association with NGC~4449.
The properties of the stream imply a massive
dwarf spheroidal progenitor, which after complete disruption will
deposit an amount of stellar mass that is
comparable to the existing stellar halo of the main galaxy.
The ratio between luminosity or stellar-mass between the
two galaxies is $\sim$~1:50, while the indirectly measured
dynamical mass-ratio, when including dark matter, may be $\sim$~1:10--1:5.
This system may thus represent a ``stealth'' merger, where an infalling satellite galaxy is
nearly undetectable by conventional means, yet has a substantial dynamical influence on its
host galaxy.
This singular discovery also suggests that satellite accretion
can play a significant role in building up the stellar halos of
low-mass galaxies, and possibly in triggering their starbursts.
\end{abstract}


\keywords{}


\section{Introduction}

A fundamental characteristic of the modern cold dark matter 
($\Lambda$CDM) cosmology \citep{2010gfe..book.....M}
is that galaxies assemble hierarchically under the
influence of gravity, continually accreting smaller DM
halos up until the present day. If those halos contain stars, then they will 
be visible as satellites
around their host galaxies, eventually disrupting through tidal forces
into distinct streams and shells before phase-mixing into obscurity
\citep{2010MNRAS.406..744C}.
This picture seems to explain the existence of satellites
and substructures observed around massive galaxies 
\citep{1966ApJS...14....1A,1988ApJ...328...88S,2009Natur.461...66M,2010AJ....140..962M}.
However, quantitative confirmation of this aspect
of $\Lambda$CDM has been more elusive,
with lingering doubts provoked by
small-scale substructure observations
\citep{2011arXiv1104.2929L,2011arXiv1111.2048B,2011arXiv1111.6609F}.

There has been relatively little work on substructure and merging
in the halos around low-mass, ``dwarf'' galaxies.
Many cases of extended stellar halos around dwarfs have been identified
observationally \citep{2009MNRAS.395.1455S}, 
but it is not clear if these stars were accreted, or formed in-situ.
Star formation in dwarfs is thought to occur in 
stochastic episodes \citep{2009ARA&A..47..371T,2011ApJ...739....5W},
which could be triggered by accretion events.

An iconic galaxy in this context is NGC~4449, a dwarf irregular in the field 
that has been studied intensively as one of the
strongest galaxy-wide starbursts in the nearby universe.
Its absolute magnitude of $M_V=-18.6$ makes it an LMC-analogue 
(and not formally a dwarf by some definitions), 
but with a much higher star formation rate.
It is strongly suspected to have recently interacted with
another galaxy based on various signatures including peculiar kinematics in its
cold gas and HII regions \citep{1986AJ.....92.1278H,1998ApJ...495L..47H},
but the nature of this interaction is unknown.

An elongated dwarf galaxy or stream candidate near NGC~4449 was first noticed by
\citet{2007AstL...33..512K} from Digitized Sky Survey (POSS-II) plates
(object d1228+4358), and is visible in the Sloan Digital Sky Survey 
(SDSS)\footnote{http://www.sdss.org/}.
Here we present deep, wide-field optical imaging
that supplies the definitive detection of this ongoing accretion event
involving a smaller galaxy, leading to 
interesting implications about the evolution of this
system and of dwarf galaxies in general.

\newpage

\section{Observations and data reduction}

Our observations of NGC~4449 and its surroundings
consist of two main components.  The first is
imaging from a small robotic telescope,
 to confirm the presence of a
low-surface-brightness substructure and provide
its basic characteristics
(similar techniques were used with larger galaxies
in \citealt{martinezDelgado2008,2010AJ....140..962M}).
The second is
follow-up imaging with the Subaru telescope to map out the resolved stellar 
populations.

We obtained very deep 
images with the f/8.3 Ritchey-Chretien 0.5-meter telescope of the 
Black Bird Remote Observatory (BBRO)\footnote{BBRO was originally
situated in the Sacramento Mountains (New Mexico, USA), and later moved to 
the Sierra 
Nevada Mountains (California, USA).} during different dark-sky observing 
runs over the periods 2010-04-13 through 2010-06-10, and
2011-01-13 through 2011-01-28 (UT).  We used a 16 mega-pixel Apogee 
Imaging Systems U16M CCD camera, with 
$31.3^\prime\times31.3^\prime$ field-of-view
and 0.46 arcsec~pix$^{-1}$ plate-scale. 
We acquired 18~hours of imaging data in half-hour sub-exposures, using a
non-infrared clear luminance 
($\lambda=$~3500--8500\AA)
Astrodon E-series filter. Each sub-exposure was reduced following standard procedures 
for dark-subtraction, bias-correction, and flat-fielding \citep{2009ApJ...692..955M}.

The resulting 
image was calibrated photometrically to SDSS using the
brighter regions of NGC~4449
(see \citealt{2010AJ....140..962M}).
The final image has 5-$\sigma$ $g$-band surface-brightness detection limits
from 26.4 to 27.5 mag~arcsec$^{-2}$ for seeing-limited and large-scale diffuse
features, respectively.

We subsequently obtained images from
the 8.2-m Subaru Telescope and the Suprime-Cam 
wide-field imager (34$^{\prime}\times$27$^\prime$ field-of-view,
$0.202^{\prime\prime}$ pixel-scale; \citealt{2002PASJ...54..833M}) on 
2011-01-05 (UT). Conditions were photometric, and we took dithered exposures in
$r^\prime$ and $i^\prime$ bands, with
total exposure times of 225 s per filter.
We reduced the data using a modified 
SDFRED pipeline \citep{2004ApJ...611..660O}, including 
bias subtraction, flat-fielding, and distortion correction.  Each frame was 
re-projected to a common astrometric coordinate system followed by 
background rectification and image co-addition using 
Montage\footnote{http://montage.ipac.caltech.edu/}.

The exquisite image quality ($\sim0.5^{\prime\prime}$ FWHM) allows us to
resolve individual stars in the outer regions of NGC~4449.
We carried out point-spread-function photometry
using DAOPHOT~II/ALLSTARS  \citep{1987PASP...99..191S},
and identified stars as objects with 
sharpness parameter $\vert{}S\vert<1.0$.
We calibrated the photometry based on two central images from
the {\it Hubble Space Telescope} Advanced Camera for Surveys (HST/ACS) 
\citep[hereafter A+08]{Ann08}.

The ACS photometry was originally in F555W and F814W, and recalibrated
to Johnson-Cousins $VI$ (A+08).
We used fairly bright, red stars in common between the datasets
to derive linear transformation equations 
from $r^\prime i^\prime$ instrumental magnitudes to $VI$,
including foreground extinction corrections of
$E(B-V)=0.019$ \citep{1998ApJ...500..525S}. 
Our final star catalog has statistical internal errors
 in $V-I$ color of $\sim$0.11 mag and $\sim$0.14 mag at $V$=25 and $V$=26, respectively.

\section{Stream morphology}

\begin{figure*}[t]
\epsfxsize=0.558\hsize
\centerline{\epsfbox{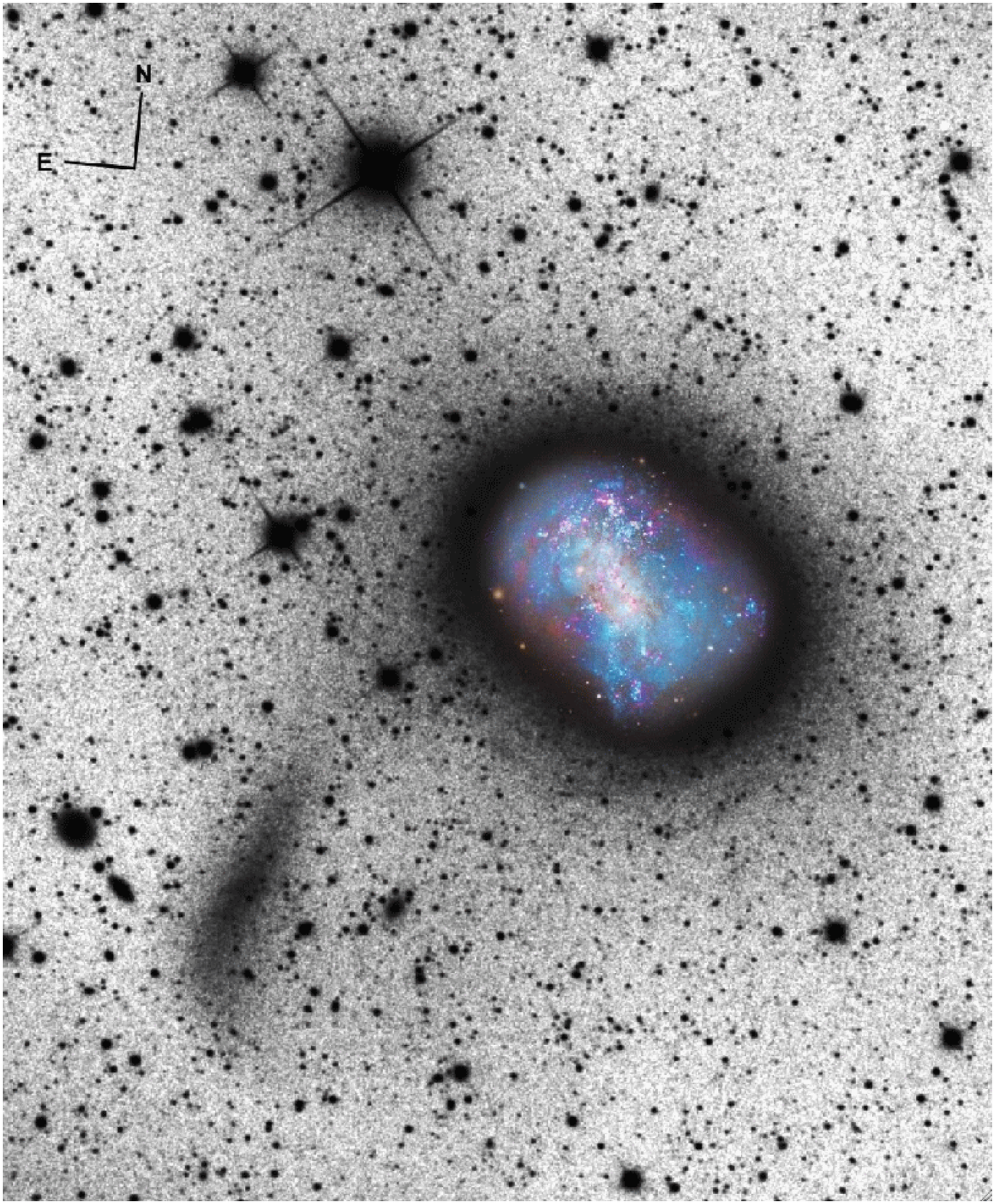}
\epsfxsize=0.442\hsize
\epsfbox{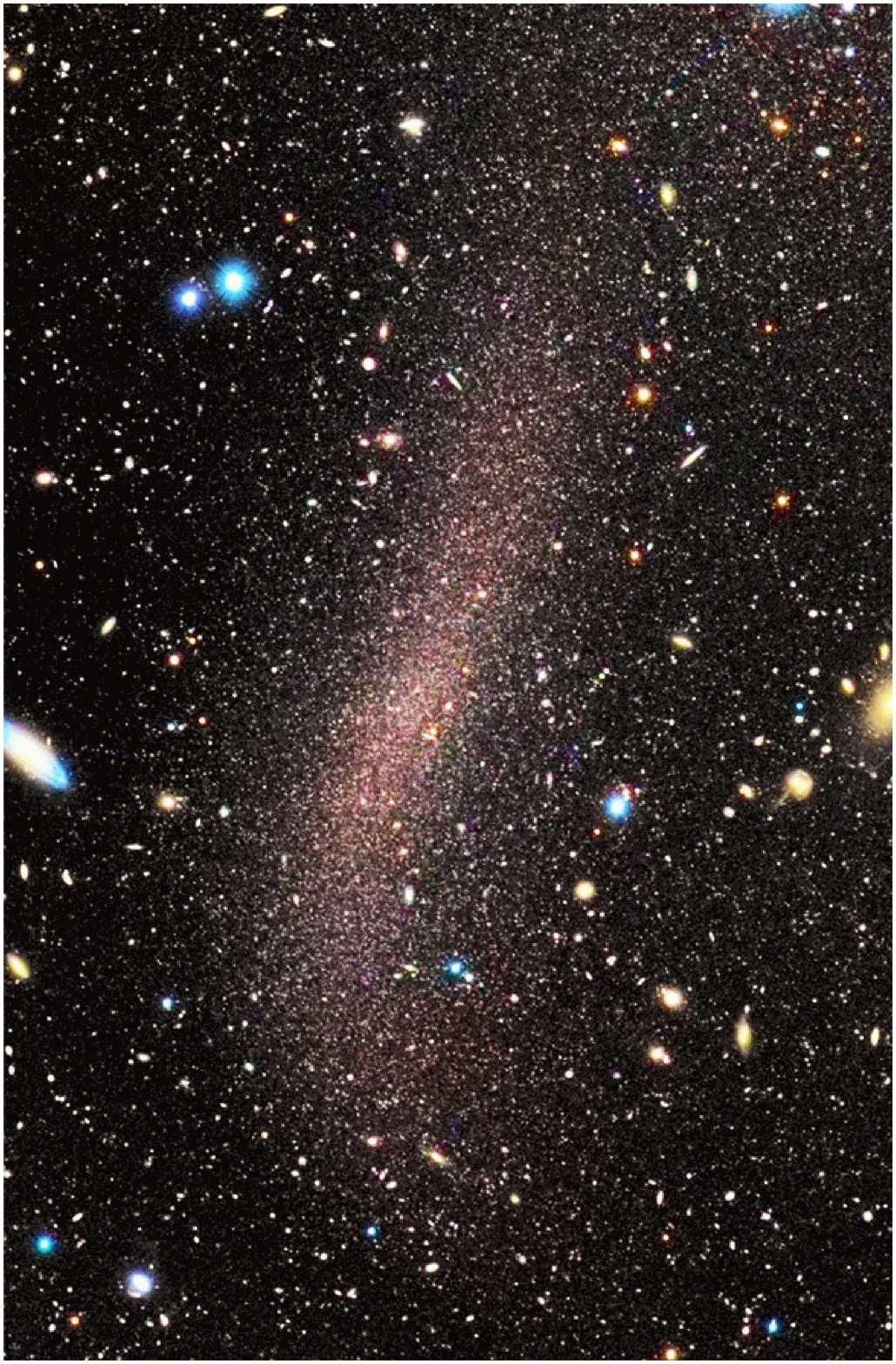}}
\figcaption{NGC~4449 and its halo stream.
{\it Left:} image from BBRO,
showing a $19.0^\prime\times24.5^\prime$ ($21\times27$~kpc) field.
{\it Right:} $5.5^\prime\times8.6^\prime$ ($6\times9.5$~kpc) subsection of the Subaru/Suprime-Cam data, showing the
 stream resolved into stars. In both panels, shallower BBRO exposures in red/green/blue filters 
provide indicative colors.
\label{f:bbo}}
\end{figure*}

Figure~\ref{f:bbo} ({\it left}) shows a BBRO image subsection, where
with an adopted distance of 3.82~Mpc (A+08),
$1^\prime$ corresponds to 1.1~kpc.
Clearly visible near the minor axis of NGC~4449,
$\sim$~10~kpc to the southeast, is
a very elongated, S-shaped feature of dimensions
$\sim 1.5\times7$~kpc, which we designate ``the stream''.
The stream's position does not overlap with any of the complex
HI-gas features surrounding NGC~4449 \citep{1998ApJ...495L..47H}.
Also, it is on the opposite side of the galaxy with respect to an interesting
star cluster that may be linked to another past accretion event \citep{2012ApJ...745L...1A}.
The main galaxy's existing stellar halo is also apparent,
including shell-like features in the southwest previously
noticed by \citet{1999AJ....118.2184H}.

The right-hand panel shows a Suprime-Cam zoom-in of the stream.
It is clearly resolved into stars
and has the general appearance of a dwarf spheroidal (dSph) galaxy
that is elongated by tides.
This classification (defined by a spheroidal morphology,
stellar mass $M_\star < 10^8 M_\odot$, and the absence of star formation and
cold gas) is also supported by the
HI non-detection of \citet{2009A&A...506..677H}.

To complement these integrated surface-brightness maps,
we construct stellar-density maps around NGC~4449 using 
individual RGB stars down to $g^\prime=25.8$ from Suprime-Cam.
We discuss the RGB selection and modeling later,
but in general these stars trace a population older than 1 Gyr.
We define a grid across the image with
$130\times130$~pc$^2$ bins,
and count the number of stars in each bin, subsequently applying
a Gaussian smoothing kernel with $\sigma_{\rm smooth}=130$~pc.
There are typically 7--9 stars per bin in the stream.

\begin{figure*}[t]
\epsfxsize=0.695\hsize
\centerline{\epsfbox{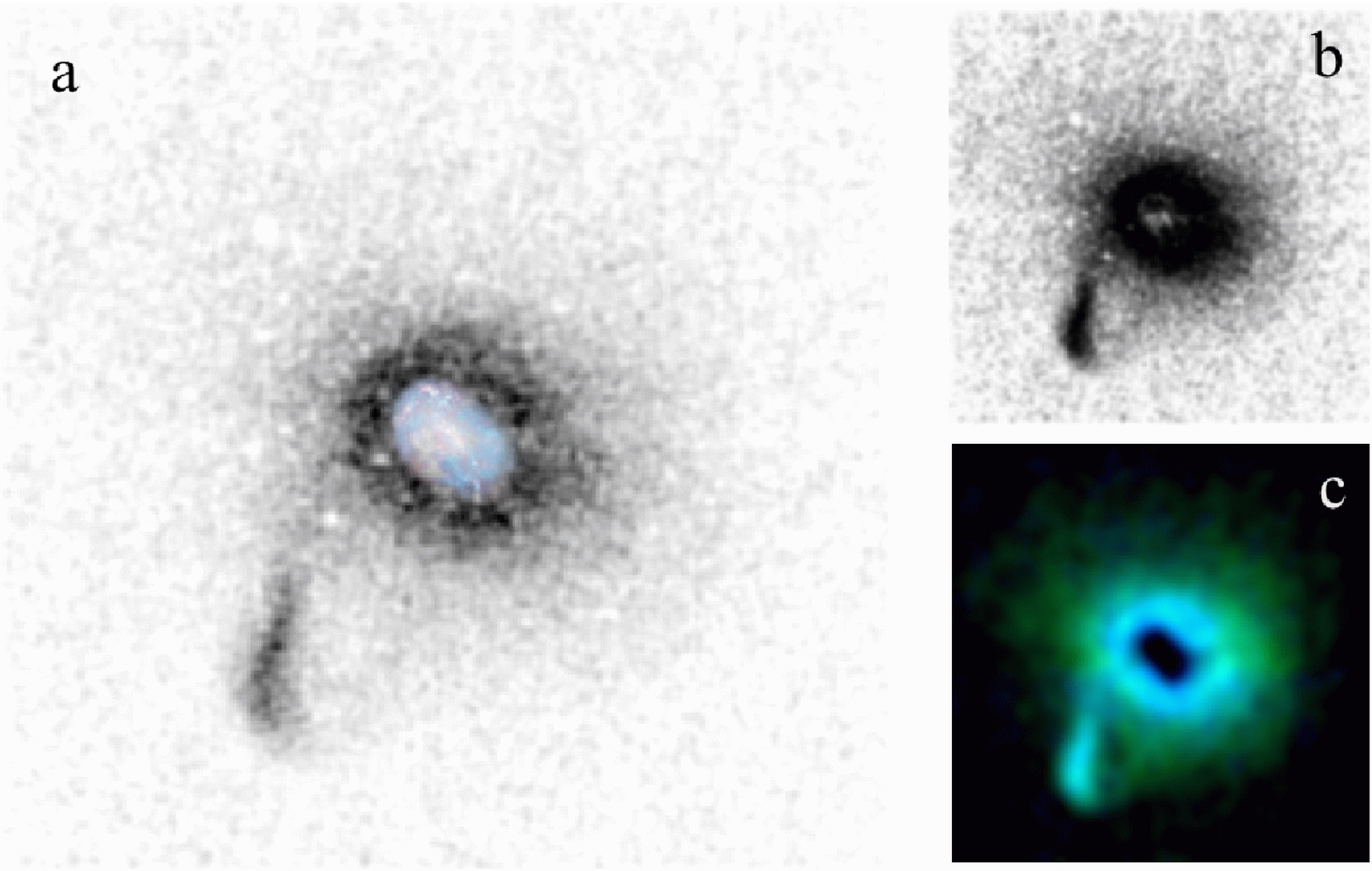}
\epsfxsize=0.305\hsize
\epsfbox{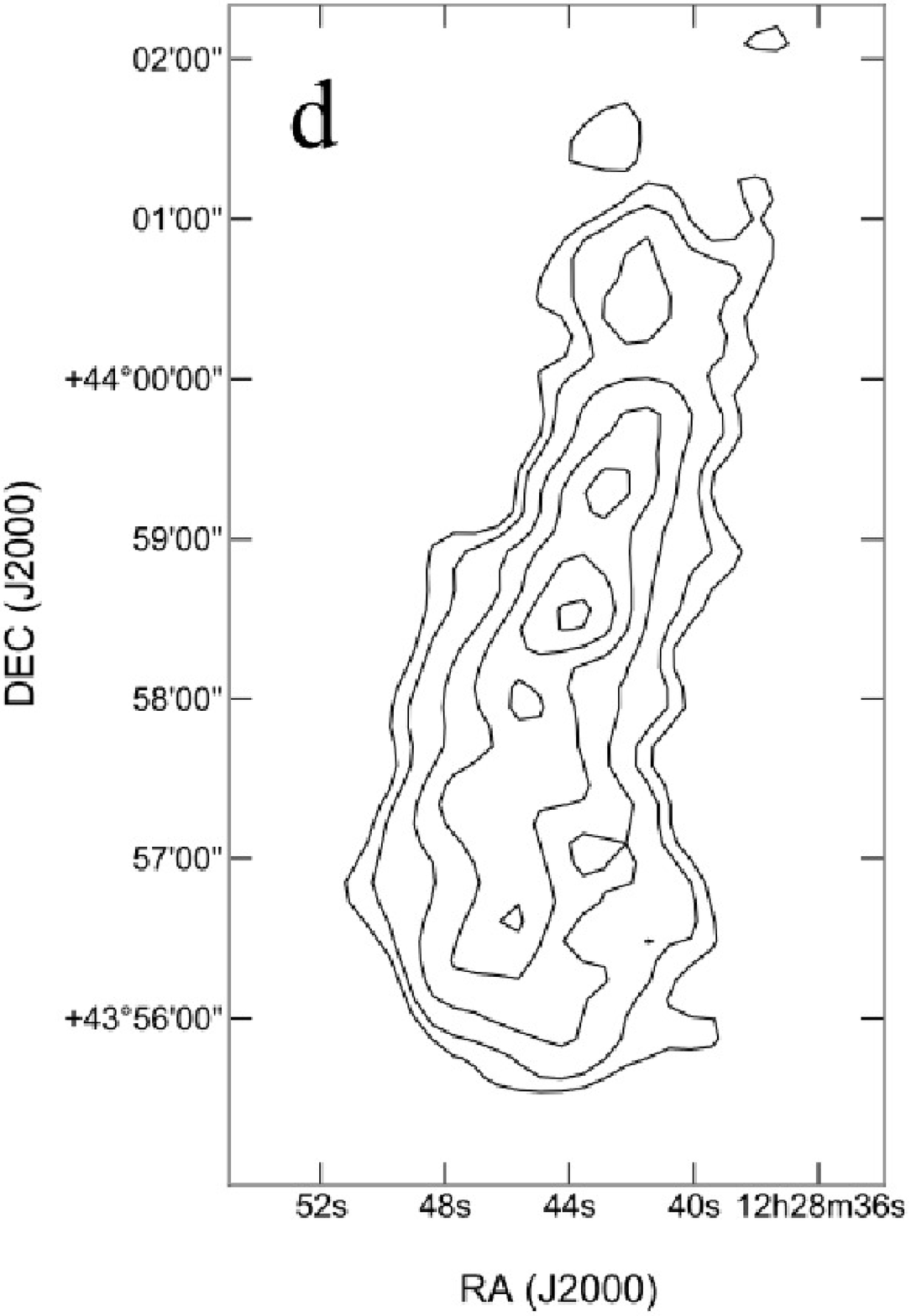}
}
\figcaption{Stellar-density maps of NGC4449 and its stream, based on RGB counts.
(a) $30^\prime\times30^\prime$ ($33\times33$~kpc) image with linear scaling.
A compact density enhancement within the stream
may be the progenitor galaxy's remnant nucleus.
(b) The same data with less dynamic range.
The stream shows a loop-like structure curving back toward
the main galaxy, which in turn also shows a shell-like overdensity
in its halo toward the southwest, reaching similar projected 
galactocentric radii as the stream.
(c) Composite map of the stream,
color-coded blue for bluer RGB colors, and green for redder colors
(see main text).
The RGBs in the stream's main part have similar average colors to the 
main galaxy's inner halo
stars, with redder populations apparent on the
outskirts of both the halo and stream.
(d) Star-count contour map of the stream,
with contour levels corresponding to 5--10 stars per 50~arcsec$^2$
bin, in intervals of $1$ star per bin.
\label{fig.map}
}
\end{figure*}

Figure~\ref{fig.map}(a) shows the resulting Subaru-based stellar-density map,
where the overall stream morphology is indistinguishable from the BBRO 
integrated-light results. Since there is
no evidence for resolved young stars (see next section), we infer that
the visible light is dominated by RGB stars 
and hence by a population older than 1 Gyr. 

Panel (b) shows a higher-contrast version of the same map, demonstrating
that the main galaxy's pre-existing stellar halo extends out to at least
$\sim$~10~kpc. We can furthermore discern a very faint feature that 
seems to extend the stream's angular path in a loop-like structure.
This loop is also apparent in the BBRO image, and
we infer that the stream is stretched out over at least half its orbit,
with the projected turning point at a galactocentric radius of 13~kpc.

We next wish to locate the disrupting satellite galaxy's original center
or nucleus. In both the BBRO and Subaru
images there is a density enhancement near the mid-point of the ``S'',
as would be expected if the ``arms'' are
leading and trailing tidal tails around a still marginally-bound,
or just disrupting, main body.
We construct a stellar-density contour map for
the stream region, using RGB stars and $\sigma_{\rm smooth}=$~270~pc.
The central stellar clump is visible in Figure~\ref{fig.map}(d),
with a position 
slightly offset from the stream's main
ridgeline
($\alpha_{\rm J2000} = 12~28~43.32, \delta_{\rm J2000} = +43~58~30.0$;
positional uncertainty $\sim 6^{\prime\prime}$).
An off-center nucleus is also seen in a tidally-disrupting
Milky Way dSph, Ursa Major \citep{2001ApJ...549L..63M}.

\section{Stellar populations}\label{sec:stellarpop}

\begin{figure*}[t]
\epsfxsize=0.6\hsize
\centerline{\epsfbox{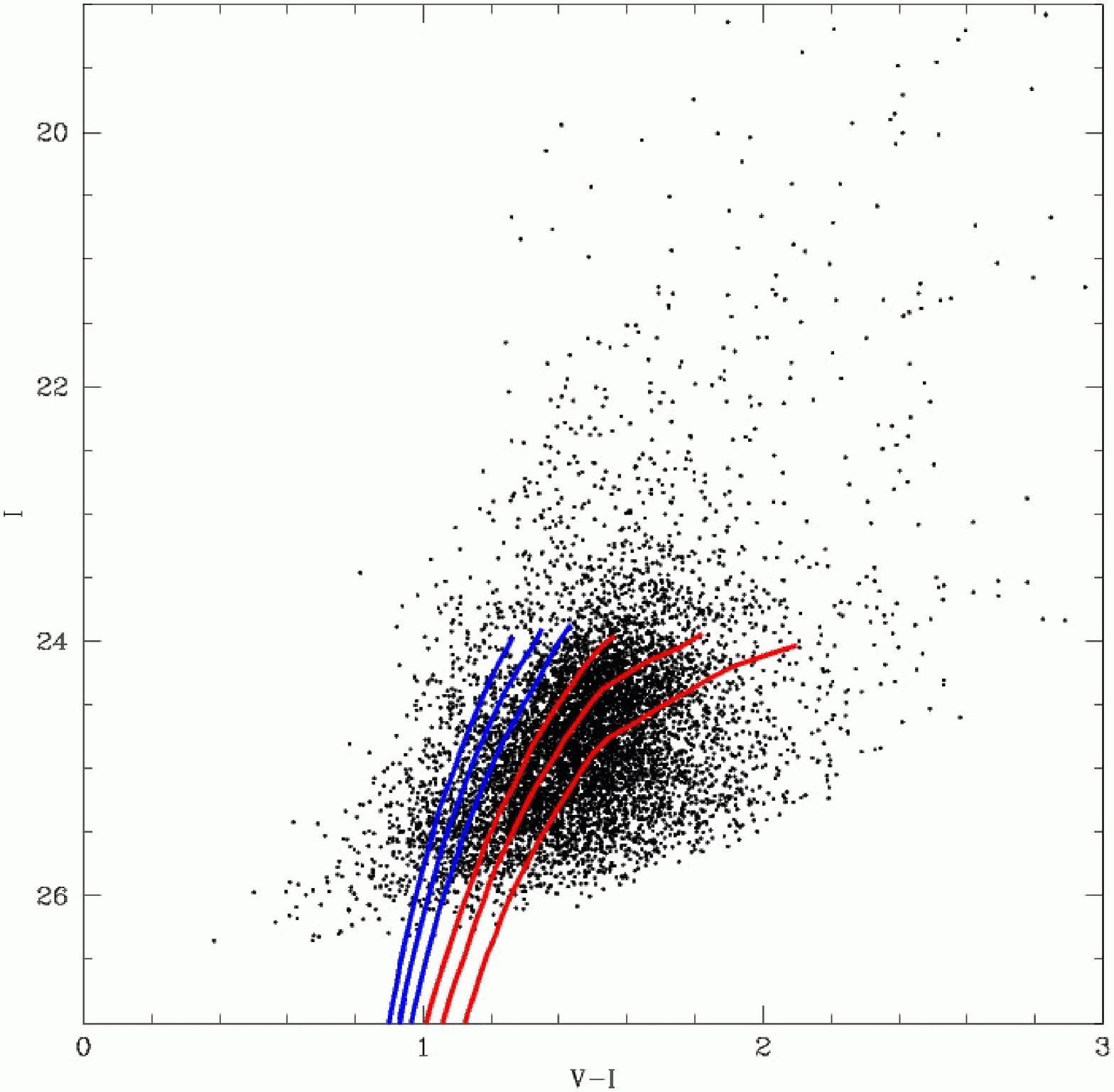}}
\figcaption{Color-magnitude diagram of the stream region, 
centered at
($\alpha_{\rm J2000}=12~28~43, \delta_{\rm J2000}=+43~58~30.0$),
with a $250^{\prime\prime} \times 540^{\prime\prime}$ ($4.6 \times 10.0$~kpc)
field-of-view, 
using $r^\prime i^\prime$ photometry transformed to $VI$. 
RGB isochrones are overlaid for metallicities
$Z=0.001$ (blue curves) and $Z=0.004$ (red curves). For each $Z$, the isochrones
show ages of 2, 4, and 10 Gyr, from left to right, respectively.\label{f:cmd}}
\end{figure*}

Figure~\ref{f:cmd} shows the color-magnitude diagram (CMD) for
point sources in the stream region.
The RGB stars are the dominant feature, along with a few brighter,
redder stars that may be oxygen- or carbon-rich
thermally-pulsating asymptotic giant branch stars from
an intermediate-age or old population \citep{Mar03}.
We find no blue stars that would trace recent star formation.

The detection of the tip of the RGB (TRGB) permits a distance estimate.
Using techniques from A+08 and \citet{Cio00}, we find
a TRGB magnitude of $I_{\rm TRGB} = 24.06 \pm 0.04$ (random) $\pm 0.08$
(systematic). The random error was estimated using bootstrapping techniques;
the systematic error is dominated by the 
magnitude-transformation uncertainties.
The main body of NGC~4449 was found by A+08 to have
$I_{\rm TRGB} = 24.00 \pm 0.01$ (random) $\pm 0.04$ (systematic). 
Thus the stream is at the same distance as the main body,
to within $\sim 180 \kpc$, and we conclude that there is a physical
association rather than a
chance superposition.

Although the RGB is affected by the well-known age-metallicity degeneracy,
this feature can still be used to constrain the properties of 
stars older than $\sim$ 1 Gyr. In Figure~\ref{f:cmd} we overplot the Padua isochrones
\citep{Gir02} for ages of $2$, $4$, and $10$ Gyr,
for both $Z=0.004$ and $Z=0.001$. For $Z=0.004$ the 2--4 Gyr 
isochrones trace the data reasonably well. The $Z=0.001$ models
appear significantly bluer than the mean RGB color
(by $\gta$~0.13 mag: our systematic color uncertainties
are $\lta$~0.1 mag), 
although a $Z=0.001$, 10 Gyr
model is consistent with the blue edge of the observed RGB. Presumably a $10$~Gyr model
with $Z\sim0.002$ would also be consistent with the data, while perhaps
providing a better match to the observed CMD slope. We conclude that, if the
bulk of the RGB stars are old ($> 10$~Gyr), the metallicity range is roughly 
$Z=0.001$--$0.004$, while for younger ages the metallicity range shifts to
higher values.

These results are comparable to the RGB analysis of the main body of
NGC~4449 by A+08 (their figure~17; see also \citealt{2011A&A...530A..23R}).
Therefore both the main galaxy and the stream contain similar
old, intermediate-metallicity populations, although the main galaxy
also contains very young stars, as well as more metal-poor
stars as inferred from its globular clusters
\citep{2012AJ....143...52S}.

We provide a preliminary overview of spatial stellar population variations
by splitting the RGBs into color-based
subpopulations, using the 4~Gyr, $Z=0.004$ model as a boundary.
We then create stellar-density maps as before, for the two subpopulations
separately.  Using blue and green color-coding to represent the
subpopulations left and right of the model boundary,
we show the results in Figure~\ref{fig.map}(c). The stream's bright parts
have no visible RGB color gradient, and have an overall color 
similar to the main galaxy's halo at radii of $\sim$~3--5~kpc.  
Both the stream's faint-loop continuation, and the halo
at $\sim$~5--10~kpc, are redder,
implying older or more metal-rich stars.

\section{Stellar and Dynamical Mass}

We now proceed to estimate the luminosities and masses of
NGC~4449 and its stream.
For NGC~4449, it is straightforward to add up the SDSS pixel fluxes
inside the ``optical radius'' ($\mu_{r}=25$~mag~arcsec$^{-2}$). 
We find an extinction-corrected $M_r=-17.8$.
For the stream, we use the SDSS-calibrated BBRO image, integrating
the flux within the faintest isophote that closes without
including the main galaxy, which corresponds to
$\mu_{g}=26.75$~mag~arcsec$^{-2}$ and a stream semi-major axis distance
of 3~kpc.
We find a stream magnitude of $M_r=-13.5$, which is
comparable to the brightest Local Group dSphs,
Fornax and And~VII.
Such galaxies have typical projected half-light radii of $\sim$~0.4--1.0~kpc
\citep{2011AJ....142..199B}, which is consistent with the stream's $\sim$~0.8~kpc half-width.

For both galaxies, these luminosity estimates are lower-limits since they
do not include potential extended low-surface-brightness features.
The implied luminosity-ratio is $\sim$~1:50.

To calculate stellar masses $M_\star$, we use two different
approaches, adopting a \citet{chabrier_03} initial-mass-function (IMF; final mass, including
stellar remnants). The first is based on the integrated optical
colors, following the relations between color and stellar
mass-to-light-ratio ($\Upsilon_\star$) from \citet[Table B1]{ZCR09}.
This paper also introduced a technique for mapping out local
$\Upsilon_\star$ and $M_\star$ variations pixel-by-pixel,
which we apply to NGC~4449, and after integrating, find
a total $M_\star=7.46\times10^8 M_\odot$.
For the stream, we assume a global color corresponding
to the central SDSS measurement, $g-r = 0.45 \pm 0.1$. We find
a stream mass of $M_\star = 1.5^{+0.8}_{-0.6} \times 10^7 M_\odot$,
implying a stellar mass-ratio between stream and host of $\sim$1:50,
for any uniform IMF.\footnote{The similarity of the luminosity- and mass-ratios
implies that the luminosity-weighted estimate of $\Upsilon_\star$
for NGC~4449 happened to turn out the same as inferred for the stream.}

The second mass-estimation approach uses the CMD,
comparing observed star-counts to predicted numbers from a
stellar populations model. For the stream, we use the $I$-band stellar
luminosity function near the TRGB, and normalize it to Monte Carlo 
simulations
drawn assuming $Z=$0.001--0.004, and ages 2--10 Gyr.
We obtain $M_\star
\sim$~(2--5.5)$\times10^7 M_\odot$ for the stream
(a lower-limit because of incompleteness),
which agrees with the color-based results.  For the main galaxy, a similar 
approach was followed by \citet{McQ10},
whose results imply $M_\star=(1.2\pm0.2)\times10^9 M_\odot$.
This mass is somewhat higher than by using colors, but the CMD-based
stellar-mass-ratio comes out to be similar, $\sim$~1:40.

Remarkably, the mass of NGC~4449's stellar halo is similar to the stream's mass:
based on the RGB counts of \citet{2011A&A...530A..23R} 
and
their normalization to $K$-band surface-brightness photometry,
we estimate $M_\star\sim2\times10^7 M_\odot$ for the halo between 
projected radii of 5--10~kpc. This halo could have therefore been built up
directly by one or a few past accretion events similar to the present-day 
stream.

We next consider the dynamical masses of the host galaxy and its 
stream, including DM. 
The quantity that is arguably the most relevant to the current stream-galaxy
interaction is the dynamical mass-ratio within the
interaction region: the $\sim$~14~kpc galactocentric radius.
Based on the HI gas kinematics, we estimate an inclination-corrected
circular velocity of $v_{\rm c}\simeq 62$~km~s$^{-1}$ at
this radius \citep{1994A&A...285..385B,2002ApJ...580..194H},
which means a dynamical mass for the main galaxy of 
$M_{\rm dyn}(r<15$~kpc)~$\simeq1.1\times10^{10}M_\odot$.
The HI gas mass is $\sim 10^9 M_\odot$ \citep{1998ApJ...495L..47H},
so this region is DM-dominated.
Note that the $v_{\rm c}$ and $M_\star$ values together suggest that
NGC~4449 is intermediate in mass to the LMC and SMC (cf. \citealt{Bes10}).

For the stream mass, we have no direct measurements, and instead
turn to a plausibility argument based on 
Local Group dSphs, where the brightest cases have estimated
circular velocities of $v_{\rm c}\sim$~15--20~km~s$^{-1}$ on $\sim$~1--3~kpc
scales \citep{2011ApJ...742...20W,2011arXiv1111.2048B}.
Then if we assume the $v_{\rm c}$ values for both stream and main galaxy
are fairly constant with radius, the ratio of $v_{\rm c}^2$ yields
the dynamical mass-ratio.
This ratio is $\sim$~1:20--1:10, and thus the stream may be significantly
perturbing the main galaxy.

A final metric is the ratio of total (virial) halo masses $M_{\rm vir}$, which
are not directly measurable but may be inferred on a statistical basis,
assuming a $\Lambda$CDM framework.  In this context, it is well-established
that the total mass-to-light-ratios of dwarf galaxies increase dramatically
at lower luminosities.  Current estimates of $M_\star$--$M_{\rm vir}$ and
luminosity--$M_{\rm vir}$ relations
\citep{2010ApJ...710..903M,2011ApJ...726..108T} 
would imply $M_{\rm
  vir}\sim$~(1--5)$\times10^{11}M_\odot$ for NGC~4449, and a pre-infall
mass of $\sim$~(1--10)$\times10^{10}M_\odot$ for the stream progenitor---which
although very uncertain, plausibly implies an
initial virial mass-ratio of $\sim$~1:10--1:5.

Thus what appears to be a very minor merger in visible light may actually
be closer to a major merger when including DM.  Such an extreme circumstance
could be compared with models of satellite disruption and potentially
discriminate between $\Lambda$CDM and alternative theories
\citep{2010ApJ...722..248M}.

\section{Discussion}

We have detected and analyzed a stellar tidal stream in the halo of NGC~4449
which we interpret as the ongoing disruption of a dSph
galaxy by a larger dwarf (an LMC/SMC analogue\footnote{The LMC and SMC may
also have a history of interaction, with a stellar mass-ratio of $\sim$~1:15
\citep{Bes10}.}).
This appears to be the lowest-mass primary galaxy with a verified stellar stream.

We suggest some implications for galaxy evolution.
It has been proposed that dSph's orbiting massive
galaxies such as the Milky Way were ``pre-processed'' from gas-rich dwarfs
by tidal effects within dwarf-galaxy groups
\citep{2009Natur.460..605D}.
We may be witnessing such a transformation in-action,
with the HI streams surrounding NGC~4449 representing additional tidal debris.

We also suspect it is not just a coincidence
that such a novel stream was found first around one of the most
intensely star-forming nearby galaxies.
The accretion event may well be the starburst trigger.
The period of elevated star formation
appears to have started $\sim$~0.5~Gyr ago \citep{McQ10},
which is suggestively similar to the stream's $\sim$~1--2~Gyr orbital
period\footnote{Given a projected apocentric radius of $R_{\rm a}=$~13~kpc,
a circular orbit provides a lower-limit for the period of
$T=2\pi R_{\rm a}/v_{\rm c}\simeq$~1.3~Gyr.}
(and to any process that is linked to the dynamical time
on $\sim$~30~kpc scales).

Are such accretion events frequent among
other dwarf galaxies in recent epochs?
We suspect that exact analogues to this stream are not very common, or they
would have been noticed already in DSS/SDSS images.
However, if the stream had been only a bit fainter, more diffuse, or
at a larger radius, it could have been missed, and thus there may be
many more dwarf-hosted stellar streams awaiting detection.

In theory, the history of DM halo assembly
should be fairly scale-free, and
$\sim$~1:10 mergers are expected to be
the most generally dominant contributors to mass growth
\citep{2008ApJ...683..597S}.
It is also increasingly recognized that such relatively minor mergers
can have important effects on the larger galaxies, such as
inciting global disk instabilities \citep{2011Natur.477..301P}.

If streams as in NGC~4449 are common in dwarfs, they re-ignite classic ideas about galaxy
interactions triggering starbursts.
Given the high rates of star formation in dwarf galaxies,
it is natural to ask if satellites are responsible.
Surveys along these lines have produced mixed results
\citep{2001A&A...371..806N,2004MNRAS.349..357B,2008MNRAS.385.1903L}, but
until now, low-surface-brightness objects such as dSphs
would have been missed.

Regardless of the implications for starbursts, dSph accretion
appears to be an increasingly viable avenue for direct assembly
of dwarf galaxies' stellar halos---as witnessed by
NGC~4449, and by Fornax, which shows traces of swallowing
an even smaller dSph \citep{2005AJ....129.1443C}.
Future observational determinations of dwarf stream frequency in combination
with theoretical models may
provide clues to the general substructure problem.

\acknowledgements

We thank Jay Strader for a preview of his paper, 
and James Bullock, Pavel Kroupa, 
Jorge Pe\~narrubia, Monica Tosi, and the referee for comments.
Based on data collected at Subaru Telescope (operated by the 
National Astronomical Observatory of Japan), via Gemini Observatory time 
exchange (GN-2010B-C-204). 
FA received partial financial support from ASI,
through contracts COFIS ASI-INAF I/016/07/0 and I/009/10/0.
The Dark Cosmology Centre is funded by the Danish National Research
Foundation. Work supported by the
National Science Foundation (Grants AST-0808099, AST-0909237,
AST-1109878, Graduate Research Fellowship), by NASA/Spitzer grant 
JPL-1310512, and by the
UCSC-UARC Aligned Research Program.




\end{document}